 \documentclass[twocolumn]{article}
\usepackage{graphicx}

\begin{document}

\title{A link between an ice age era and a rapid polar shift} 

\author{W. Woelfli\footnote{ Institute for Particle Physics, ETHZ H\"onggerberg, CH-8093 Z\"urich, Switzerland (Prof.~emerit.); e-mail: woelfli@phys.ethz.ch .} \ and
W. Baltensperger\footnote{Centro Brasileiro de Pesquisas F\'\i sicas, Rua Dr.\thinspace Xavier Sigaud,150, 222\thinspace 90 Rio de Janeiro, Brazil;\qquad \qquad e-mail: baltens@cbpf.br}}
 % \date{\small (Version 19a, edited  12/07/2004,\quad typeset \the\day /\the\month /\the\year )}
\maketitle

{\noindent\emph{The striking asymmetry of the ice cover during the Last Global Maximum suggests that the North Pole was in Greenland and then rapidly shifted to its present position in the Arctic See. A scenario which causes such a rapid geographic polar shift is physically possible. It involves an additional planet, which disappeared by evaporation within the Holocene.  This is only possible within such a short period, if the planet was in an extremely eccentric orbit and hot. Then, since this produced an interplanetary gas cloud, the polar shift had to be preceded by a cold period with large global temperature variations during several million years. }} 

\vspace{0.5cm}
In the Last Global Maximum the ice cover on the North American continent reached the region of present New York and in Europe of North Germany, while East Arctic Siberia was ice free. This well known asymmetry is corroborated by current research. The palaeolithic Yana RHS site \cite{Pitulko} at 71$^\circ$N, well inside the polar circle, was populated 27\thinspace 000 years ago. This is referred to as "A surprising survival story" \cite{Stone}. Alternatively, the site is an indication that the climate in that region was mild. Clearly, the Sun was not warmer since on the North American continent the climate was cold.  In arctic Siberia, a mild climate could only result from lower latitude. 

The postulate of a geographic polar shift is more than a century old. The striking asymmetry of the ice cover on the northern hemisphere in the Last Global Maximum would disappear, if  the north pole had then been situated within Greenland, shifted by about 18$^\circ$ from its present position in the Arctic Ocean \cite{Woelfli2002}. Considering this shift, the Yana RHS site would have a position of about 53$^\circ$N, appropriate for hunting and living all year round. Of course, other regions also  experienced changes. For instance in the Amazon palaeo sand dunes have been found at the Equator \cite{Amazon}. Before the shift, these were at a latitude of about 18$^\circ$N, which crosses the present Sahara desert.  The assumed turn of the globe similarly explains climate changes of many regions.  Since the South Pole remained within Antarctica, the resulting climate changes on the southern hemisphere were less spectacular. 

	How can a polar shift arise? The shape of the rotating Earth is determined by hydrostatic equilibrium; the Earth has a slight bulge along the Equator. For this reason, the rotation axis perpendicular to the equatorial plane is stable. If, suddenly, the Earth got a further deformation in some oblique direction, the stable position of the rotation axis would be in a different direction. Then, geographically, the present rotation axis would start to move around this stable direction. This motion would continue until the Earth had completely relaxed into a new hydrostatic equilibrium shape. During the polar wandering, Earth's angular momentum is strictly conserved.  Since the Earth is nearly a sphere the rotation axis makes a small angle of less than a minute of arc with the angular momentum vector so that its direction approximately also remains pointing to a fixed star. It is the globe itself that moves. The path of the rotation axis on the globe is a spiral. The difference between end point and starting point is the geographic polar shift. A turn in the spiral takes about 400 days (Chandler period). For simplicity, we shall describe the decay of the difference between Earth's actual shape and its equilibrium shape for the instantaneous rotation axis by a relaxation time. Then, for a significant shift, the relaxation time must be longer than about 100 days. Otherwise the motion stops too soon. This required slow relaxation characterizes a plastic behavior, a concept of the $20^{th}$  century which was missing in early treatments of the problem. A solid would relax with the speed of sound and a liquid with the speed of liquid flow. Since matter needs to be displaced only about 10 km, the equilibrium shape of Earth would be reached within less than one day in these cases. Thomas Gold \cite{Gold} called attention to the fact that with a slow relaxation (of, say, 1000 days for global deformations) a geographic polar shift becomes possible. Numerically, the motion is obtained from two equations. First, a relaxation equation of the inertial tensor describes the gradual replacement of the old equatorial bulge and additional deformation by a new equatorial bulge belonging to the instantaneous rotation axis. Second, the equation of motion of the gyroscope, the Euler equation, determines the path of the rotation axis in coordinates fixed to the Earth. Admittedly, this is a simplified treatment of the problem. Most statements of this paper are based on simple estimates. More detailed models may change numbers or perhaps even conclusions. The required shift of 18$^\circ$ can be obtained by a stretching deformation with an amplitude of 1 per mil in a direction 30$^\circ$ to the initial rotation axis assuming a relaxation time of 1000 days. Note that the maximum change of latitude which is larger than the final change, is reached in 200 days. The violent variation of climate evidenced by frozen mammoths thus becomes  plausible. 

What can cause this assumed sudden additional deformation?  An asteroid hitting the Earth would introduce a lot of energy and momentum into one point. This is not suitable for a soft displacement of large quantities of matter. We need a more globally acting force.

Let us suppose that an additional planet, henceforth called Z, passes near the Earth. Its gravitational attraction accelerates the Earth. At Earth's center, inertia compensates the perturbing gravitational force. In the part of the Earth near Z, the gravitational attraction dominates, while on the far side, it is the inertia. Therefore this tidal force stretches the globe along the direction to Z. If the closest approach between the centers of Earth and of Z is 20\thinspace 000 km and their relative velocity 40 km/s, then the passage lasts for about 500 s $\approx 8$ minutes. In this time a stretching amplitude of 6.5 km must be built up. It is easy to estimate the equilibrium deformation for a constant tidal force \cite{Woelfli1999}. The dynamic problem is far more complicated. We simply suppose that the peak tidal force then has to be an order of magnitude larger. With this assumption and  giving Z the mass of Mars, the closest approach between the centers of Z and Earth is about 15\thinspace 000 km. A close approach indeed! This shows that the mass of Z cannot be smaller than that of Mars. Obviously, a larger planet passing at a greater distance would have more time and could produce the deformation in a smoother way. 

This planet Z does not exist any more. How can it have disappeared in the short time span of the Holocene? Only the Sun can achieve this by either evaporating or swallowing Z. Every possibility must be considered to this end. The relative velocity between Earth and Z is a planetary velocity determined by the mass of the Sun. The momentum transfer during the close approach amounts to only about 1 \% of the momentum of Z, so that its orbit changes little. Therefore, already before the approach, the orbit of Z must have been extremely eccentric.  Indeed, Z had to approach the Sun as close as compatible with its existence, say to a perihelion distance of about 4 million km. Near the perihelion Z was heated by solar radiation and tidal work, thus was liquid and shining. A highly eccentric orbit has a small angular momentum. This can suffer a relative change of order $\pm 10$ \% during the close approach between Z and Earth, and the resulting relative change of the perihelion distance is twice as big. The disappearance of Z during the short Holocene indicates that the angular momentum of Z's orbit decreased in the actual event. Nevertheless, a direct drop into the Sun after the close approach seems unlikely.
 
During the close approach the tidal forces on Z were larger than those on Earth. Z may have been torn into two or more pieces (called Z$_n$).  The fact that Z was liquid and hot inside facilitated its disintegration. A disruption of Z implies that the mass of Z was smaller than Earth's since otherwise the Earth would have been broken up too. This restricts the possible mass of Z to a range of one order of magnitude.

	Since Z was hot it lost matter by evaporation. For a Mars sized planet this is limited by the escape energy of the particles. Atoms or ions evaporate rather than molecules. The evaporation rate from the pieces Z$_n$ is dramatically increased. Continued tidal work between Sun and the parts Z$_n$ causes their perihelion distances to shrink. This, combined with a continuously diminishing escape velocity, may lead to the complete evaporation of the objects Z$_n$ within the Holocene.

Atoms and ions (from Z$_n$  also molecules) evaporated into interplanetary space. Particles are not blown away by radiation pressure provided their first excitation energy lies beyond the main solar spectrum. Atoms and ions of this type remained in orbits bound to the Sun. Yet, even these particles scatter part of the solar radiation (mostly Thomson scattering, visible as violet-blue). This limits their lifetime; due to Poynting-Robertson drag \cite{Dermott} they spiral into the Sun. While Z was hot, a cloud of particles existed in interplanetary space, which screened part of the solar radiation. It is an observed fact \cite{Tiedemann} that about 3 million years ago the mean temperature of Earth started to decrease and to fluctuate. We conclude that Z must have been in the highly eccentric orbit since that time. In reality and in this model the era of glaciation has a beginning.

 	The distance of approach necessary for a polar shift, say 20\thinspace 000 km, defines a target which is minute compared to the surface of the sphere at Earth's distance from the Sun (150 million km). Therefore for a polar shift event Z probably crossed this sphere millions of times, even assuming that the plane of its orbit  remained within a few degrees from the ecliptic. On the other hand, the trial period could not have lasted much longer since otherwise Z would have collided with one of the inner planets considering  that the distance of approach for a polar shift is not much larger than the sum of the radii relevant for a collision. To sum up, the occurrence of the polar shift together with the fact that Z disappeared in less than 10\thinspace 000 years require that Z had been in an extremely eccentric orbit during several million years. During this time the gas cloud affected the climates of the planets. 

This cloud of atoms and ions in interplanetary space probably has complex properties. Two particles in planetary orbits typically have a sufficiently high relative velocitiy so that their collision is inelastic and accompanied by light emission. This reduces their relative kinetic energy. The orbits become more similar. The width of the cloud shrinks. This in turn shortens the time between collisions. The dynamics of the cloud furthermore depends on the lifetime of the particles. If the Earth is in the cloud, part of the solar radiation is screened and Earth's temperature globally reduced. (Actually, if the Earth were in an isotropically distributed cloud, the incidence of radiation would be enhanced by backscattering.) However, if the cloud is  restricted to a narrow solid angle, the Earth -- being outside the cloud -- is exposed to the full radiation from most of the solar surface and in addition some scattered radiation from the cloud. In this situation the climate can be globally warmer than it is at present \cite{Ravelo}. When the Earth is in the cloud, atoms and ions flow into Earth's atmosphere. This is clearly demonstrated by the sharply peaked measured occurrence of impurities in the ice of Antarctica during the cold periods \cite{EPICA}. During the last three million years the mass of Z decreased due to evaporation and the perihelion distance of Z shrank because of tidal work. Both effects enhanced the evaporation rate. Thus the temperature of the cold periods decreased with time and its variations became larger \cite{Tiedemann,Ravelo,EPICA}. The time dependence of the screening depends on the external forcing on Earth and on the particles. We expect that this, in conjunction with the feedbacks on Earth, will form a basis for understanding the observed temperature variations, in particular the Dansgaard-Oeschger \cite{GRIP} events. At present, the basic distribution of climates on Earth follows plausible rules. Therefore, the surprising climate variations during the ice age era are the result of a violent action from the outside on the intrinsically well behaved climate system with displaced poles. The inclination, i.e. the angle between Earth's orbit and the invariant plane (perpendicular to the solar system's angular momentum), should be an important parameter for the screening of the solar radiation. Since Earth's inclination has a period of 100 kyr, this can be the key for the observed dominant temperature variations of the last 700 kyr \cite{EPICA,Muller95,Muller97,Woelfli02}.
 
The origin of the postulated planet Z is not known. It may have been a moon of Jupiter or it may have come from outside the planetary system. This ice age era with its temperature variations is a strange and possibly unique period in Earth's history. Correspondingly, the occurrence of a planet like Z has to be extremely rare. Z presumably lost energy through resonances with Jupiter and other planets in a time of the order of one million years \cite{Dermott,Murray}. The fact that it simultaneously lost angular momentum and entered an extremely eccentric orbit is not exceptional \cite{Murray}.  There are not many adjustable parameters in the model. The mass of  Z prior the end of the Pleistocene is between 1/10 of Earth's mass and Earth's mass. The perihelion distance is such that Z is hot during the Pleistocene. We assume 4 Mio km but this number depends on the treatment of the heating of Z by irradiation and tidal work. The major semi-axis and eccentricity of Z's orbit must combine to an aphelion distance that is larger than the radius of Earth's orbit. However, a much larger value of the major semi-axis would inhibit sufficiently frequent interactions with the Earth \cite{Nufer}. An extreme eccentricity makes the orbit sensitive to loss of angular momentum  during the close encounter and thereby helps Z$_n$ to disappear during the Holocene. Further orbital parameters which would define Z's position at given time are not determined. We can only discuss typical behaviours of an orbiting Z \cite{Nufer}. 

Striking mysteries of palaeoclimate can be explained by a geographic polar shift near the end of the Pleistocene.  Therefore, there have been numerous attempts \cite{Hapgood} to find a mechanism for such a shift. This paper describes one of these. At the present stage, the paper tries to call attention to the fact that a shift of the geographic position of the poles is physically possible. This is a catastrophic event as evidenced by frozen animals and muck in arctic Siberia. Their C$^{14}$ ages mostly scatter around 11\thinspace 500 years before present. It is an open question, why some frozen carcasses have different C$^{14}$ ages.
 
In this model the polar shift results from the close passage of an additional planet in our planetary system. A polar shift scenario in which this planet is in a moderately eccentric orbit could also be considered. In this case there would be no gas cloud and no ice age era. The additional planet could be expelled from the solar system or driven into the Sun by resonances in a time of the order of a million years. However, the condition that Z must disappear within the Holocene limits the scenarios to a narrow range. In this case Z must be in an extremely eccentric orbit so that it is hot and evaporates. It then follows that the polar shift must be preceded by a cold period with vigorous temperature fluctuations during a few million years. The ice age era which lasted for about 3 million years \cite{Tiedemann} is such a period. The model explains that it had a beginning and end. With the rapid disappearance of Z, the ice age era is an inevitable precursor of the pole shift event.

\end{document}